\documentclass[]{elsarticle}
\usepackage{graphicx}
\usepackage{amsmath}
\usepackage{natbib}
\usepackage{bm}
\usepackage{xcolor}

\newcommand{\vvec}[1]{\mathbf{#1}}
\newcommand{\mathbb}[1]{\mathbf{#1}}

\date{\today}

\begin{document}

\title{On the swelling properties of  pom-pom polymers in dilute solutions. Part 1: symmetric case}
\author{Khristine Haydukivska\corref{cor1}}
\ead{wja4eslawa@gmail.com}
\author{Ostap Kalyuzhnyi}
\author{Viktoria Blavatska}
\author{Jaroslav Ilnytskyi}
\address{Institute for Condensed Matter Physics of the National Academy of  Sciences of Ukraine,1, Svientsitskii Str., 79011 Lviv, Ukraine}
\cortext[cor1]{Corresponding author}

\begin{abstract}
We consider the simplest representative of the class of  multiply branched polymer macromolecules, known as a pom-pom structure. The molecule consists of a backbone linear chain terminated by two branching points with functionalities (numbers of side chains) of $f_1$ and $f_2$, respectively. In a symmetrical case, considered in the present study, one has $f_1=f_2=f$ with the total number of chains $F=2f+1$. Whereas rheological behaviour of melts of pom-pom molecules are intensively studied so far, we turn our attention towards conformational properties of such polymers in a regime of dilute solution. The universality concept, originated in the critical phenomena and in scaling properties of polymers, is used in this study. To be able to compare the outcome of the direct polymer renormalization approach with that obtained via dissipative particle dynamics simulations, we concentrated on the universal ratios of the shape characteristics. In this way the differences in the energy and length scales are eliminated and the universal ratios depend only on space dimension, solvent quality and a type of molecular branching. Such universal ratios were evaluated both for a whole molecule and for its individual branches. For some shape properties, theory and simulations are in excellent agreement, for other we found the interval of $F$ where both agree reasonably well. Combination of theoretical and simulation approaches provide thorough quantitative description of the peculiarities of swelling effects and spatial extension of pom-pom molecules and are compared with the known results for simpler molecular topologies.
\end{abstract}


\maketitle

\section{Introduction}

Prediction of the effects of branching on the properties of macromolecules in solution is a problem
of continuing interest in polymer physics since the pioneering work
of Zimm and Stockmayer in 1949 \cite{Zimm49}.
The simplest representative of this class of polymers is the so-called star polymer architecture
 with a number of linear branches radiating from a single branching point, which is thoroughly studied by now \cite{star}.
However, from industrial point of view  the studies of
macromolecules of more complex structure are very important,
 since regular star type branching is not the only one that makes its way into commercial polymers.
 The multiply- and hyper-branched polymers attract a lot of attention
due to the high processability of such objects: their properties can be tuned by controlling number,
length, and distribution of branches along the backbone chain.

It is established, that polymer melts of macromolecules with multiple branching sites
 have rheological properties that differ distinctly
from those of linear or short-branched star polymers due to effects
of entanglement between molecules \cite{Meissner72,Mcleish95}.
Typical example are commercial polymer melts, such as low density
polyethylene (LDPE), which consists of linear polyethylene backbones
with attached alkyl branches. The multiple long-chain branches
influence the processing properties considerably and results in
lowering the viscosity and occurrence of strain-hardening
phenomenon in uniaxial extensional flow \cite{McLeish98}.

\begin{figure}[h!]
\begin{center}
\includegraphics[width=73mm]{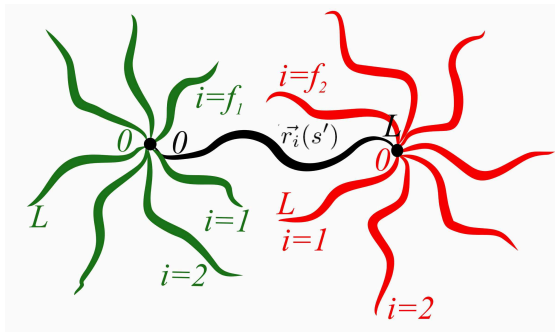}
\caption{ \label{fig:1} Schematic presentation of a pom-pom polymer structure. The backbone linear chain has two branching points at both of its ends,
with functionalities $f_1$
and $f_2$, respectively. }
\end{center}
\end{figure}

In this concern, different kinds of model branched polymer
structures can be considered, e.g. the H-shape
\cite{Roovers84,McLeish88} or more general comb-shape
\cite{Roovers75,Roovers79,Lipson87,Wang93,Redke05} architectures.
The simplest model of polymer architecture that correctly captures the
nonlinear rheological characteristics of commercial long-chain
branched polymers, coined the term of a pom-pom structure
\cite{Bishko97}. This class of branched polymers can be considered
as a generalization of H-polymer structure. The molecule consists of
a backbone linear chain with two branching points
of functionalities $f_1$ and $f_2$ at its both ends (see Fig.
\ref{fig:1}).
 For simplicity, in what follows we will consider the case when the lengths of the radiating branches and backbone
 are equal:
$L_i=L$, $i=1,\ldots,F$, { where  $F\equiv f_1+f_2+1$ denotes the total number of linear strands in pom-pom topology}. Note, that historically such a structure
is mentioned for the first time in the work of Zimm and Stockmayer
\cite{Zimm49}, as a ``chain with two branch units'', whereas in some
other papers (like Ref. \citenum{Bayer94}) such architecture is called
``dumbbell model''. Introduction of this model was a breakthrough in
the field of viscoelastic constitutive equations. Linear and
nonlinear rheology of model branched polymers with pom-pom
architecture  was investigated in
Refs. \cite{McLeish98,Graham01,Ruymbeke07,Chen11}. The series of
highly entangled polystyrenes with pom-pom architecture
synthesized anionically
\cite{Nielsen06}. Recently, the macromolecules of poly(lactic acid)
(PLA), one of the most successful environment-friendly commercial
polymers, were synthesized  \cite{Gu17} both in star-shape and
pom-pom-like shapes to compare the properties of these topologies.

Whereas the viscoelastic and dynamical properties of dense
melts of multiply branched polymers are of principal importance from
industrial and technological points of view, the conformational
properties of such macromolecules in a very diluted regime also
attract considerable interest. Good example of practical applications of dilute polymer solutions is their use as lubricants. Here molecular topology of a polymer plays an important role, e.g. the use of star polymers allows not only to modify viscosity of a solution but also to improve certain other properties of lubricants\cite{Shell}. It is a well established fact that branched polymers are characterised by lower intrinsic viscosity than their linear counterparts of the same mass. In practice this fact is quantified by introducing the viscosity shrinking factor which is equal to a ratio between the viscosities of the branched and liner polymers\cite{Khabaz14,knauss2002,Fer2019}. The experimental values for the viscosity shrinking factor for the pom-pom polymers indicate that this ratio decreases with the increase of the level of polymer branching\cite{knauss2002}. This is traditionally related to the fact that branched polymers are more compact in size as compared to their linear counterparts, and the effective polymer size is related directly to the intrinsic viscosity through the Flory-Fox equation\cite{Berry2007,Kok1981}.
As suggested in Ref. \citenum{Zimm49}, the level of molecular ``compactization'' due to the presence of branching can be characterised by the ratio between the gyration radius of branched molecule and that of the equivalent mass liner chain. This ratio has also a practical application as it is directly related to the respective change in the intrinsic viscosity of the solution\cite{Fer2019,Zimm1959}. The size ratio of the pom-pom structure and a linear chain of the equivalent total mass in simplest case of Gaussian polymers without monomer-monomer excluded volume is given by \cite{Zimm49,Radke96}:
\begin{equation}
g_{f_1,f_2}=\frac{3(f_1^2+f_2^2)+4(f_1+f_2)+12f_1f_2+1}{(f_1+f_2+1)^2}. \label{gpom}
\end{equation}
{
Note that in the rest of the paper we will be interested in simplified case of so-called symmetric pom-pom molecules
with $f_1=f_2=f$, so that
\begin{equation}
g_{f,f}=\frac{18f^2+8f+1}{(2f+1)^2}. \label{gpomsym}
\end{equation}}
For the case of star architecture with one branching point of functionality $f$ the corresponding size ratio reads:
\begin{equation}\label{gf}
g(f)=\frac{3f-2}{f^2}.
\end{equation}
These are examples of universal characteristics in conformational
properties of macromolecu\-les, which are independent on any details
of chemical structure and are governed only by so-called global
parameters. In Gaussian case, the functionalities of branching points serve
as global parameters. Dimensional dependence of the size ratio
$g(f)$ is found by introducing the concept of excluded volume.
Analytical~\cite{Miyake84,Alessandrini92} and
numerical~\cite{Batoulis89,Bishop93a,Bishop93b,Wei97} studies
indicate the increase of value $g(f)$ in this case as comparing with
Gaussian polymers. The size ratios of H-comb polymers (which can be
treated as symmetric pom-pom structure with $f=2$) in
presence of excluded volume have been estimated numerically in Refs.
\cite{Bishop92,Bishop93a,Bishop93b,Khabaz14} and analytically in Refs.
\cite{Kosmas89,Douglas84}.

More subtle universal characteristics, specific to bran\-ched polymers,
are the branch stretch ratio $p_e(f)$ defined as ratio of the averaged center-end distance  of individual branch of star polymer to the end-to-end distance of independent linear chain of the same
molecular weight,  and the branch swelling ratio $p_g(f)$ given by ratio of gyration radii of individual branch of a star and that of linear chain.
 These ratios characterize the averaged effect of crowdedness caused by a presence of all other branches  on a size of a given individual branch.
 For the case of star-branched polymers, these values have been thoroughly analyzed both analytically and numerically in our previous study \cite{Kalyuzhnyi2019}.

 Besides the effect of compactization, branching also drastically changes the shape characteristics of polymer macromolecules.  It is convenient to characterise the asymmetry of polymer configurations in
terms of rotationally invariant universal quantities
\cite{Aronovitz86,Rudnick86} constructed as combinations of the
components of the gyration tensor, such as the asphericity $A_d$.
This quantity takes on a maximum value of unity for a completely
stretched configuration, and equals zero for the spherical form. In
general, to characterize the level of ``spherization'' of the
branched polymer, the ratio of asphericities of branched stracture
and the linear chain of the total molecular weight can be considered
  \cite{Batoulis89}.
The shape characteristics of H-polymers have been studied
numerically in \cite{Zweier09,Ferber15}.

Whereas size and shape characteristics of star-bran\-ched and H-polymers have been thoroughly analyzed so far via both
numerical and analytical approaches, much less attention had been paid in this concern to pom-pom architecture.
 The  aim of our study is to contribute into the statistical description of conformation properties of macromolecules of symmetric pom-pom topology in dilute solutions. To this end, we apply both the analytical approach, based on continuous polymer model and renormalization scheme, and numerical mesoscopic dissipative particle dynamics based simulations. Both methods allow us to obtain quantitative estimates for the set of universal size and shape characteristics of interest.

The layout of the rest of the paper is as follows. The next Section 2 \ref{Theory} is dedicated to analytical analysis of the problem of interest. We introduce the continious model presentation of pom-pom molecule, shortly describe the direct polymer renormalization approach and present our results up to the first order of perturbation theory expansion. In the following Section 3 \ref{Numeric} we cover the mesoscopic simulation approach of dissipative particle dynamics and provide our numerical results obtained by it.
We end up by giving conclusions and outlook.


\section{Theoretical approach}\label{Theory}
\subsection{Continuous chain model} \label{M}

In the frames of continuous chain model \cite{Edwards},  the pom-pom structure as presented in
Fig. \ref{fig:1} is considered as a set of $F$  trajectories of
length $L$,  parametrized by radius vectors $\vvec{r}_i(s)$, $0\leq s \leq L$, $i=1,\ldots,F$.
 Hamiltonian of such structure reads:
\begin{eqnarray}
&&H_F = \frac{1}{2}\sum_{i=1}^{F}\,\int_0^{L} ds\,\left(\frac{d\vvec{r_i}(s)}{ds}\right)^2+\frac{u}{2}\sum_{i,j=1}^{F}\int_0^{L}ds'\int_0^{L} ds''\,\delta(\vvec{r_i}(s')-\vvec{r_j}(s'')),\label{H}
\end{eqnarray}
where the first term describes the chain connectivity and the second one corresponds to excluded volume interaction with coupling constant $u$.
Note that, by performing the dimensional analysis of couplings in above expression, we find that $[u]=[L]^{(4-d)/2}$. The ``upper critical'' value
of space dimension $d_c=4$, at which the coupling becomes dimensionless, is important in developing the renormalization scheme, as will be shown in next section.

The topology itself is taken into account when introducing the partition function for the special topology.
For the star-like structure with $f_s$ branches it can be determined as:
\begin{eqnarray}
Z^{{\rm star}}_{f_s}=\frac{1}{Z_0^{{\rm star}}}\int\,D\vvec{r}(s)\,\prod_{i=1}^{f_s}\delta(\vvec{r_i}(0))\,{\rm e}^{-H_{f_s}},
\label{Zs}
\end{eqnarray}
where the product of $\delta$-functions describes the fact that all trajectories $\vvec{r_i}$, $i=1,\ldots,f_s$
 start at the same point and $Z_0$ is a partition function of a Gaussian molecule is:
 \begin{eqnarray}
&& Z_0^{{\rm star}}=\int\,D\vvec{r}(s)\,\prod_{i=1}^{f_s}\delta(\vvec{r_i}(0))\,{\rm e}^{-\frac{1}{2}\sum_{i=1}^{f_s}\,\int_0^L ds\,\left(\frac{d\vvec{r_i}(s)}{ds}\right)^2}.
 \end{eqnarray}
The partition function for pom-pom  structure is given by:
\begin{eqnarray}
&&Z^{{\rm pom-pom}}_{f_1,f_2}=\frac{1}{Z_0^{{\rm pom-pom}}}\int\,D\vvec{r}(s)
\prod_{i=1}^{f_1}\prod_{j=1}^{f_2}\,\delta(\vvec{r_i}(0)-\vvec{r_0}(0))\nonumber\\
&&\delta(\vvec{r_j}(0)-\vvec{r_0}(L))\,{\rm e}^{-H},
\label{ZZ}
\end{eqnarray}
where we count the main backbone chain as $0$th, and a set of $f_1$ and  $f_2$ trajectories start at its origin ($\vvec{r_0}(0)$)
and end point ($\vvec{r_0}(L)$), correspondingly. The partition function of corresponding Gaussian molecule reads
 \begin{eqnarray}
 &&Z_0^{{\rm pom-pom}}=\int\,D\vvec{r}(s)\,\prod_{i=1}^{f_1}\prod_{j=1}^{f_2}\,\delta(\vvec{r_i}(0)-\vvec{r_0}(0))\nonumber\\
 &&\times\delta(\vvec{r_j}(0)-\vvec{r_0}(L))\,{\rm e}^{-\frac{1}{2}\sum_{i=0}^{f_1+f_2+1}\,\int_0^{L} ds\,\left(\frac{d\vvec{r_i}(s)}{ds}\right)^2
 }.
 \end{eqnarray}

\subsection{Direct renormalization method} \label{Met}

On the base of continuous chain model the observables are expressed as perturbation theory series in coupling constant $u$, which diverge in the limit of $L\rightarrow \infty$. These divergencies need to be eliminated
to receive the universal values of the parameters under consideration. For this purpose the direct renormalizatin method was developed by des Cloiseaux \cite{desCloiseaux}. The key point of the method is to introduce the
renormalization factors that are directly connected to the observable physical quantities and
allow to remove the divergences. In order to receive the finite values of observables one needs to evaluate them at the corresponding
fixed points (FPs) of the renormalization group.

The  FPs  values do not depend on the topology of the macromolecule, and as a result can be calculated in the simplest case of single linear chain. The task of calculating the FP starts from  obtaining the partition function of two interacting polymers $Z(L,L)$. It is also important to introduce the renormalization factors $[Z(L,u_0)]^{-2}$, with $Z$ being the partition function of a single chain and $\chi_0(L,\{x_0\})$ --  a so-called swelling factor  connected to end-to-end distance $\chi_0=\langle R_e^2\rangle/L$. A renormalized coupling constant can be presented as:
\begin{eqnarray}
u_R(u_0)= - [Z(L,u_0)]^{-2}Z(L,L)[2\pi \chi_0(L,u_0)]^{-2+\epsilon/2},
\end{eqnarray}
where $\epsilon=4-d$ is a deviation from upper critical dimension for the coupling constant $u_0$. Taking a limit of infinitely long chain in this expression one can receive the fixed values of coupling constants:
\begin{eqnarray}
\lim_{L\to\infty} u_{R}(u_0)=u_R^*.
\end{eqnarray}
In the first order of perturbation theory one has \cite{desCloiseaux}:
\begin{eqnarray}
&& {\rm {Gaussian}}: u^*_{R}=0,\qquad {\text at}\qquad d\geq4,\label{FPG}\\
&& { \rm {Pure}}: \qquad u^*_{R}=\frac{\epsilon}{8}.\qquad {\text at}\qquad d<4.\label{FPP}
\end{eqnarray}
Here, Eq. (\ref{FPG}) corresponds to the idealized Gaussian molecule without any interactions
between monomers, whereas Eq. (\ref{FPP}) describes the influence of excluded volume effect.

\subsection{The Douglas-Freed approximation for the size ratios}\label{approximation}
It is known that a reasonably good agreement between the results obtained via renormalization group approach and the experiment is achieved when at least the second order terms in coupling constant $u_0$ are taken into account. In some cases this requires performing massive calculations. Continuous chain model can be related to the two parameter model at $d=3$, and then the approximation can be used within a renormalization group approach developed by Douglas and Freed \cite{Douglas84} that allows for better match between renormalisation group in the first order with the experimental data.

Within the renormalization group approach any observable can be presented in a form of universal function. The expression for gyration radius  reads
\begin{equation}
\langle R_g^2 \rangle = \langle R_g^2 \rangle_0 \left(\frac{2\pi N}{\Lambda}\right)^{2\nu(\eta)-1}f_p(\eta),\label{SF}
\end{equation}
with $N$ being the number of monomers, $\Lambda$ is a coarse-graining length scale,
and $f_p(\eta)$ is a function that  equals  1 for $\eta=0$ (Gaussian chain) and $1+a$ when $\eta\rightarrow \infty$ which corresponds to  the case of good solvent and $a$ is topology dependent. Equation (\ref{SF}) has the same form for any polymer topology, and as a result the size ratio of gyration radii of macromolecules of two topologies can be presented as:
\begin{equation}
g_x = \frac{\langle R_{g,1}^2 \rangle_0}{\langle R_{g,2}^2 \rangle_0}\frac{1+a_1}{1+a_2}.
\end{equation}
In the frames of the renormalization group approach, parameters $a$ can be obtained when deriving (\ref{SF}) for a particular topology \cite{Douglas84} considering at least terms of the order of $\epsilon^2$. However, when that is impossible, it was suggested to use the results from the two-parameter model for which the gyration radius can be presented in the first order of perturbation parameter $z$ as:
\begin{equation}
\langle R_g^2 \rangle = \langle R_g^2 \rangle_0 \left(1+Cz\right),\label{df}
\end{equation}
with $C$ being the the function of branching parameters calculated in three dimensional space related to the parameter $a$ for gyration radius via:
 \begin{equation}
a=\frac{3}{32}C-\frac{1}{4}.
\end{equation}

\subsection{Results}

\subsubsection{Partition function}
\begin{figure}[b!]
	\begin{center}
		\includegraphics[width=73mm]{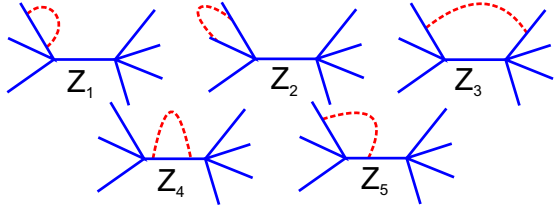}
		\caption{ \label{fig:2}Diagrammatic presentations of the contributions to the partition function up to the first order of perturbation theory in coupling constant $u$.
			The solid lines are schematic presentations of polymer strands and dash line represents a two monomer excluded volume interaction.}
	\end{center}
\end{figure}

We start with considering the partition function of a pom-pom polymer which is determined by expression (\ref{ZZ}).
Following the general scheme as given in Ref. \citenum{desCloiseaux}, and generalized to more complex polymer topologies in previous works \cite{Blavatska12,Blavatska15}, we evaluate
it as a perturbation theory series in an excluded volume coupling constant $u$.
We restrict ourself to the first order of expansion in $u$ and a symmetrical case of $f_1=f_2=f$.
In order to calculate the corresponding expression,  it is useful to
exploit the diagrammatic technique (see Fig. \ref{fig:2}). The corresponding combinatorial factors should be prescribed to each diagram.
The diagram $Z_1$ corresponds to all the excluded volume interactions within one chain, it  should be taken
with  pre-factor $2f+1$. Diagram $Z_2$ denotes the contributions from all the interactions between two chains that are separated by one branching point and thus contain a pre-factor $2f+2f(f-1)$. Diagram $Z_3$ corresponds to interactions
between chains separated by two branching points and thus has a pre-factor $f^2$. Diagram $Z_4$ appears only once, while $Z_5$ is to be taken with combinatorial factor $2f$.
Corresponding analytical expressions read:
\begin{eqnarray}
&&Z_1=\frac{u(2\pi)^{-d/2}L^{2-d/2}}{(1-d/2)(2-d/2)},\\
&&Z_2=\frac{u(2\pi)^{-d/2}L^{2-d/2}(2^{2-d/2}-2)}{(1-d/2)(2-d/2)},\\
&&Z_3=\frac{u(2\pi)^{-d/2}L^{2-d/2}\left(3^{2-d/2}-2^{3-d/2}+1\right)}{(1-d/2)(2-d/2)}.
\end{eqnarray}
Evaluating the contributions as
series in  deviation from the upper critical dimension $\epsilon=4-d$, we receive:
\begin{eqnarray}
&&Z_1=u_0\left(-\frac{2}{\epsilon}-1\right),\\
&&Z_2=u_0\left(\frac{2}{\epsilon}+1-\ln(2)\right),\\
&&Z_3=u_0\left(2\ln(2)-\ln(3)\right),
\end{eqnarray}
where we made the reduced coupling constant is $u_0=u(2\pi)^{-d/2}L^{2-d/2}$.

As a result, we obtain the expression for the partition function of pom-pom structure in one-loop approximation:
\begin{eqnarray}
&&Z^{{\rm pom-pom}}_{f,f}=1-u_0\left(\frac{2f(f-1)-2}{\epsilon}{-}2f{-}1\right.+\nonumber\\
&&\left.+(2f+f(f-1))(1-\ln(2))+f^2(2\ln(2)-\ln(3))\right). \label{Zfinal}
\end{eqnarray}

In what follows, the partition function (\ref{Zfinal}) will be used in evaluation of the averaged values of observables under interest,
defined by:
\begin{eqnarray}
&&\langle (\ldots) \rangle = \frac{1}{{ Z^{pom-pom}_{f,f}}}\prod_{i=1}^{f}\prod_{j=1}^{f}\,\int\,D\vvec{r}(s)\,\times\nonumber\\
&&\times\delta(\vvec{r_i}(0)-\vvec{r_0}(0))\delta(\vvec{r_j}(0)-\vvec{r_0}(L))\,{\rm e}^{-H}(\ldots).
\end{eqnarray}

\subsubsection{Size characteristics}

To describe the size properties of the pom-pom polymer structure we will concentrate attention on a number of characteristics,
which can be to some extend related to previous research on the subject.

{\bf End-to-end distances in pom-pom structure.}
We start with considering the end-to-end distances in pom-pom structure, that can be useful to characterise the spatial
extensions of molecule in solvent. Here we are considering:

1) distance between the free end of randomly chosen branch (side chain) in one of the ``pom'' structure and its host branching point,
thus defined as:
\begin{eqnarray}
{\langle {r^2_{{\rm branch}}}\rangle} = \frac{1}{f}\sum_{i=1}^{f}{\langle(\vvec{r}_i(L)-\vvec{r}_i(0))^2\rangle};
\end{eqnarray}
2) the end-to-end distance for the backbone:
\begin{eqnarray}
{\langle {r^2_{{\rm backbone}}}\rangle} = {\langle(\vvec{r}_0(L)-\vvec{r}_0(0))^2\rangle};
\end{eqnarray}
3) distance between the free ends of two randomly chosen branches of the same ``pom'':
\begin{eqnarray}
{\langle {r^2_{{\rm f_1,f_1}}}\rangle} = \frac{2}{f(f-1)}\sum_{i,j=1}^{f}{\langle(\vvec{r}_i(L)+\vvec{r}_j(L))^2\rangle};\label{rperp}
\end{eqnarray}
4) distance between the free ends of two randomly chosen branches of different ``poms'':
\begin{eqnarray}
{\langle {r^2_{{\rm f_1,f_2}}}\rangle} = \frac{1}{f^2}\sum_{i,j=1}^{f}{\langle(\vvec{r}_i(L)+\vvec{r}_0(L)+\vvec{r}_j(L))^2\rangle}.\label{rpar}
\end{eqnarray}

 To find the corresponding analytical expression for the first quantity, we use the identity:
\begin{eqnarray}
&&\langle(\vvec{r}_i(L)-\vvec{r}_i(0))^2\rangle = - 2 \frac{d}{d|\vvec{k}|^2}\xi_e(\vvec{k})_{\vvec{k}=0},\nonumber\\
&&\xi_e(\vvec{k})\equiv\langle{\rm e}^{-\iota\vvec{k}(\vvec{r}_i(L)-\vvec{r}_i(0))}\rangle,
\end{eqnarray}
where $\iota$ is an complex unity.
Exploiting the same scheme, as for partition function in previous section, we obtain:
\begin{eqnarray}
&&\langle r_{branch}^2\rangle=Ld\left(1+u\left(\frac{2}{\epsilon}-1-\frac{f}{3}-2f(\ln(2)-\ln(3))\right)\right).\label{e_branch}
\end{eqnarray}
In the similar way can be defined the corresponding identities for the remaining quantities and the expressions will read:
\begin{eqnarray}
&&\langle r_{backbone}^2\rangle=Ld\left(1+u\left(\frac{2}{\epsilon}-1+2f\ln(2)-\frac{f}{2}+\frac{f^2}{6} \label{e_backbone}
\right)\right),\\
&&\langle r_{f_1,f_1}^2\rangle=2Ld\left(1+u\left(\frac{2}{\epsilon}-\frac{4f+9}{12} -2f(\ln(2)-\ln(3))\label{e_pom} \right)\right),\\
&&\langle r_{f_1,f_2}^2\rangle=3Ld\left(1+u\left(\frac{2}{\epsilon}+\frac {\ln(2)(2f-2)}{3} +\frac {\ln(3)(1+2f)}{3}\right.\right.\nonumber\\&&\left.\left.-\frac {f^2-9f-10}{18}\right)\right).\label{e_pom-pom}
\end{eqnarray}

{\bf Gyration radii}

\begin{figure}[t!]
	\begin{center}
		\includegraphics[width=73mm]{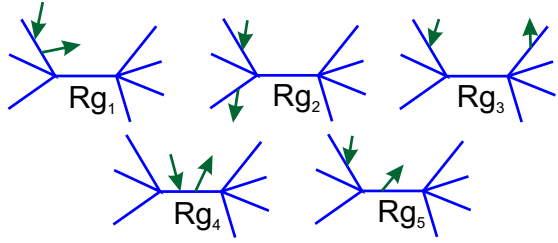}
		\caption{ \label{fig:3}Diagrammatic presentation of contribution into $\xi(\vvec{k})$ in Gaussian approximation. The solid lines are schematic presentations of polymer strands each of the length $L$ and arrows represent the so-called restriction points $s_1$ and $s_2$.}
	\end{center}
\end{figure}
Within the continuous chain model the gyration radiuses for the pom-pom polymer structure can be presented as:
\begin{eqnarray}
&&{\langle {R^2_{g}}\rangle} = \frac{1}{2L^2(F)^2}  \sum_{i,j=0}^{F}\int_0^{L}\int_0^{L} ds_1\,ds_2 \langle(\vvec{r}_i(s_2)-\vvec{r}_j(s_1))^2\rangle, \label{totalRg}
\end{eqnarray}
where $F$ is a total number of branches for which the gyration radius is calculated. Again,
 we can consider different special cases:

1) radius of gyration of a single branch in a ``pom'', averaged over all linear strands in the structure
\begin{eqnarray}
&&{\langle {r^2_{g, branch}}\rangle} = \frac{1}{2L^2(F-1)^2}  \sum_{i=1}^{F}\int_0^{L}\int_0^{L} ds_1\,ds_2 \langle(\vvec{r}_i(s_2)-\vvec{r}_i(s_1))^2\rangle;
\end{eqnarray}

2) radius of gyration of a backbone
\begin{eqnarray}
&&{\langle {r^2_{g, backbone}}\rangle} = \frac{1}{L^2} \int_0^{L}\int_0^{L} ds_1\,ds_2 \langle(\vvec{r}_0(s_2)-\vvec{r}_0(s_1))^2\rangle;
\end{eqnarray}

3) radius of  gyration of one ``pom''
\begin{eqnarray}
&&{\langle {r^2_{g, pom}}\rangle} = \frac{1}{2L^2f^2}  \sum_{i=1}^{f}\int_0^{L}\int_0^{L} ds_1\,ds_2 \langle(\vvec{r}_i(s_2)-\vvec{r}_i(s_1))^2\rangle;
\end{eqnarray}

4) radius of gyration of the full structure ${\langle {R^2_{g, pom-pom}}\rangle}$, given by Eq. (\ref{totalRg}).

It is useful to present the identiny:
\begin{eqnarray}
&&\langle(\vvec{r}_i(s_2)-\vvec{r}_j(s_1))^2\rangle = - 2 \frac{d}{d|\vvec{k}|^2}\xi(\vvec{k})_{\vvec{k}=0},\nonumber\\
&&\xi(\vvec{k})\equiv\langle{\rm e}^{-\iota\vvec{k}(\vvec{r}_i(s_2)-\vvec{r}_j(s_1))}\rangle.
\end{eqnarray}
and evaluate $\xi(\vvec{k})$ in path integration approach.
In calculations of the contributions to $\xi(\vvec{k})$, it is convenient to use the diagrammatic presentation, as given in
Fig. \ref{fig:3} for the Gaussian chain.
It is interesting to note that diagrams in Figs. \ref{fig:2} and \ref{fig:3} look similar with only difference that in first case we consider
them with interaction points and in second case with the restriction points, however the pre-factor will be the same.


In the case of a single branch in one of the ``poms'' we have:
\begin{eqnarray}
&&\langle r^2_{g,\,{\rm branch}}\rangle=\frac{dL}{6}\left(1+ \frac{2u_0}{\epsilon}+u_0\left(-\frac{13}{12}+\frac{89}{6}f+36f(\ln(2)-\ln(3))\right)\right),\label{g_branch}
\end{eqnarray}
while the expression for a backbone reads:	
\begin{eqnarray}
&&\langle r^2_{g,\,{\rm backbone}}\rangle=\frac{dL}{6}\left(1+ \frac{2u_0}{\epsilon}+u_0\left(-\frac{13}{12}+\frac{35 f}{4}+\frac {f^2}{12}-12f\ln(2)\right)\right).\label{g_backbone}
\end{eqnarray}
Combined gyration radiuses of all the branches on one side of of a single ``pom'' is given by the expression:
\begin{eqnarray}
&&\langle r^2_{g,\,{\rm pom}}\rangle=\langle R^2_{g,star}\rangle_0\left(1+ \frac{2u_0}{\epsilon}+\frac{u_0}{12(3f-2)}\right.\nonumber\\
&&\times\left(864(\ln(2)-\ln(3))f^2-1680\ln(2)f+\right.\nonumber\\
&&\left.+1728\ln(3)f+304f^2-48\ln(2)-633f-1\right).\label{g_pom}
\end{eqnarray}
Note that the prefactor $\langle R^2_{g,star}\rangle_0=\frac{dL(3f-2)}{6f}$ in this expression reproduces the known value of gyration radius of individual star polymer.

Finally, for the gyration radius of the whole structure we obtained an expression:
\begin{eqnarray}
&&\langle R^2_{g,pom-pom}\rangle=\langle R^2_{g,pom-pom}\rangle_0\left(1+u_0\left(\frac{2}{\epsilon} \right.\right.\nonumber\\
&&+\frac{1}{12(6f+1)}\left(12f^4+1296 f^3\ln(3)\right.\nonumber\\
&&-864f^3\ln(2)-972\ln(3)f^2+1056\ln(2)f^2\nonumber\\
&&\left.\left.\left.-648f^3-192\ln(2)f+246f^2+52f-13 \right)\right)\right)\label{g_pom-pom}
\end{eqnarray}
with prefactor giving the corresponding value of Gaussian structure
\begin{eqnarray}
&&\langle R^2_{g,pom-pom}\rangle_0 =\frac{dL}{6(2f+1)^2}\left(3f^2{+}\frac{4}{3}f{+}\frac{1}{6}\right).\label{Rg0}
\end{eqnarray}
{Note that here we present results for the $\epsilon$-expansion whereas expressions as the functions of the space dimension are provided in the appendix.

\subsubsection{Universal size ratios}

Within the continuous chain model all the parameters are calculated in the limit of infinitely long chains thus in order to compare results of calculations with ether experimental data or results from numerical simulations it is traditional \cite{Douglas84,Zimm49,Blavatska12} to consider a size ratio between the structure of interest (in this case pom-pom one) and a simpler one. Such ratios are universal (see \ref{approximation}), thus do not depend on the length scale of the molecule and allow to quantitatively compare two structures with different topologies. In this work we consider two of such ratios:
\begin{eqnarray}
&&g_{s}=\frac{\langle R^2_{g,star}\rangle}{\langle R^2_{g,pom-pom}\rangle},\label{gs}\\
&&g_{c}=\frac{\langle R^2_{g,pom-pom}\rangle}{\langle R^2_{g,chain}\rangle},\label{gc}
\end{eqnarray}
where $\langle R^2_{g,star}\rangle$ is a gyration radius of a star polymer of the same total molecular weight $F$ \cite{Blavatska12}:
\begin{eqnarray}
&&\langle R^2_{g,\,{\rm star}}\rangle = \frac{dL(12F^2+8F+1)}{6(2F+1)}
\left(1+ \frac{2u_0}{\epsilon}\right.+\frac{u_0}{12(6F+1)}\left(576\ln(2)F^2\right.\nonumber  \\
&&\left.\left.-192\ln(2)F-312F^2+78F-13\right) \right) \label{Rstar}
\end{eqnarray}
and  $\langle R^2_{g,chain}\rangle$ is gyration radius of chain of length $F$\cite{desCloiseaux}  :
\begin{eqnarray}
\langle R^2_{g,\,{\rm chain}}\rangle= \frac{dL}{6(F L)}\left(1+ \frac{2u_0}{\epsilon}+u_0\ln(FL)-\frac{13}{12}\right).
\end{eqnarray}
 We obtained the following estimates:
\begin{eqnarray}
&&g_s=\frac{12f^2+8f+1}{18f^2+8f+1}\left(1{-}\frac{u_0f^2}{(6f+1)(18f^2+8f+1)}\left(648\ln(3)f^2-1296\ln(2)f^2\right.\right.\nonumber\\
&&\left.\left.+6f^3-378\ln(3)f+(360f+72)\ln(2)+145f^2-81\ln(3)+160f+40\right)\right),\label{gs1}\\
&&g_c=\frac{18f^2+8f+1}{(2f+1)^3}\left(1-u_0\left(\frac{27\ln(3)f^2(3-4f)}{18f^2+8f+1}
+\frac{f(1-f)(f^2-53f-13)}{18f^2+8f+1}\right.\right.\nonumber\\
&&\left.\left.+\frac{8\ln(2)f(9f-2)(f-1)}{18f^2+8f+1}+\ln(1+2f)\right)\right).\label{gc1}
\end{eqnarray}
We evaluate the expressions (\ref{gs1}) and (\ref{gc1}) at fixed point (\ref{FPP}) for the case of three dimensional space ($\epsilon=1$).

Another interesting quantity is the ratio of the end-to end distances of free ends of a branch
on one ``pom'' (defined by Eq. (\ref{rperp}) ) versus that on different ``poms'' (Eq. (\ref{rpar})):
\begin{eqnarray}
&&g_{e}=\frac{\langle r_{f_1,f_1}^2\rangle}{\langle r_{f_1,f_2}^2\rangle},\label{ge}
\end{eqnarray}
which can characterise so-called ``acylindricity'' of pom-pom structure.
 Using expressions  (\ref{e_pom}) and (\ref{e_pom-pom}), we evaluate the  ratio:
\begin{eqnarray}
&&g_e=\frac{2}{3}\left(1+u_0\left(\frac{1}{3}(4f-1)(\ln(3)-2\ln(2)) -\frac{7}{36}+\frac{f}{6}-\frac{f^2}{18}\right)\right). \label{ge1}
\end{eqnarray}
Again, we evaluate it at substituting the fixed point value (\ref{FPP}) for the case of three dimensional space ($\epsilon=1$).

Another set of stretching and swelling ratios are introduced to  describe the effects of crowdedness caused by presence of many entangled
polymer strands of pom-pom structure on the properties of its individual branches.  As it was mentioned above, we distinguish between the backbone strand and side chains, which  are expected to demonstrate different stretching behavior.
To illustrate this we consider a set of stretch and swelling ratios:
\begin{eqnarray}
\label{pca} &&p_{e,c/a}=\frac{\langle r_{backbone}^2\rangle}{\langle r^2_{branch}\rangle},\quad p_{g,c/a}=\frac{\langle r_{g,backbone}^2\rangle}{\langle r^2_{g,branch}\rangle},\\
\label{pccch} &&p_{e,c/ch}=\frac{\langle r_{backbone}^2\rangle}{\langle r^2_{chain}\rangle},\quad p_{g,c/ch}=\frac{\langle r_{g,backbone}^2\rangle}{\langle r^2_{g, chain}\rangle},\\
\label{pach} &&p_{e,a/ch}=\frac{\langle r_{branch}^2\rangle}{\langle
	r^2_{chain}\rangle},\quad p_{g,a/ch}=\frac{\langle
	r_{g,branch}^2\rangle}{\langle r^2_{g,chain}\rangle}.
\end{eqnarray}
In analytical evaluation  up to the first order of perturbation theory we obtain:
\begin{eqnarray}
&&p_{e,c/a}=1+u_0\left(\frac{f}{6}(24\ln(2)-1+f
-12\ln(3))\right)\label{peca_res},\\
&&p_{e,c/ch}=1-u_0\left(\frac{f}{2}-\frac{f^2}{6}-2f\ln(2)\right)\label{pecch_res},\\
&&p_{e,a/ch}=1-u_0\left(\frac{f}{3}+2f\left(\ln(2)-\ln(3)\right)\right)\label{peach_res},\\
&&p_{g,c/a}=1-\frac{f u_0}{12}\left(73-f+576\ln(2)-432\ln(3)\right)\label{pgca_res},\\
&&p_{g,c/ch}=1-u_0\left(12f\ln(2)-\frac{35f}{4}-\frac{f^2}{12}\right)\label{pgcch_res},\\
&&p_{g,a/ch}=1+\frac{u_0 f}{6}\left(89+216\ln(2)-216\ln(3)\right)\label{pgach_res}.
\end{eqnarray}
The quantitative estimates of above expressions as functions of branching parameter can be obtained by substituting the fixed point values (\ref{FPP}) for the case of three dimensional space ($\epsilon=1$). However those numbers will allow for only qualitative comparison with results received in simulation. The better quantitative comparison for the one loop approximation of the ratios $g_c$ and $g_s$ can be reached by using the approach proposed by Douglas and Freed \cite{Douglas84}. Results for other ratios can not be determined through this approximation as they are much more sensitive to the excluded volume interaction and for the quantitative comparison with the results of simulation a higher orders of perturbation theory are required.

The results for all the ratios are discussed in details in the next section, alongside with their comparison with the computer simulation data.

\section{Simulation approach}\label{Numeric}

\subsection{Dissipative particle dynamics simulations} \label{IV}

In this paper we are using the dissipative particle dynamics (DPD), a mesoscopic simulation technique
following closely Ref.~\citenum{Groot1997}. In this
approach the polymer and solvent molecules are modeled as soft beads
of equal size, each representing a group of atoms. The length-scale is
given by the diameter of a soft bead, and the energy unit is
$k_{B}T$, where $k_{B}$ is the Boltzmann constant and $T$ is the
temperature. Monomers in a polymer chain are bonded via harmonic
springs resulting in a force:
\begin{equation}\label{FB}
\vvec{F}^B_{ij} = -k\vvec{x}_{ij}\,,
\end{equation}
where $k$ is the spring constant, and
$\vvec{x}_{ij}=\vvec{x}_i-\vvec{x}_j$, $\vvec{x}_i$ and $\vvec{x}_j$
are the coordinates of $i$th and $j$th bead, respectively. The
non-bonding force $\vvec{F}_{ij}$ acting on the $i$th bead as the result of its interaction with
its $j$th counterpart is expressed as a sum of three contributions
\begin{equation}
\vvec{F}_{ij} = \vvec{F}^{\mathrm{C}}_{ij} + \vvec{F}^{\mathrm{D}}_{ij}
+ \vvec{F}^{\mathrm{R}}_{ij}\,,
\end{equation}
where $\vvec{F}^{\mathrm{C}}_{ij}$ is the conservative force responsible for the repulsion between the beads,
$\vvec{F}^{\mathrm{D}}_{ij}$ is the dissipative force mimicking the friction between them and the random force
$\vvec{F}^{\mathrm{R}}_{ij}$  works in pair with a dissipative force
to thermostat the system. The expression for all these three
contribution are given below \cite{Groot1997}
\begin{equation}\label{FC}
\vvec{F}^{\mathrm{C}}_{ij} =
\left\{
\begin{array}{ll}
a(1-x_{ij})\displaystyle\frac{\vvec{x}_{ij}}{x_{ij}}, & x_{ij}<1,\\
0,                       & x_{ij}\geq 1,
\end{array}
\right.
\end{equation}
\begin{equation}\label{FD}
\vvec{F}^{\mathrm{D}}_{ij} = -\gamma
w^{\mathrm{D}}(x_{ij})(\vvec{x}_{ij}\cdot\vvec{v}_{ij})\frac{\vvec{x}_{ij}}{x^2_{ij}},
\end{equation}
\begin{equation}\label{FR}
\vvec{F}^{\mathrm{R}}_{ij} = \sigma
w^{\mathrm{R}}(x_{ij})\theta_{ij}\Delta t^{-1/2}\frac{\vvec{x}_{ij}}{x_{ij}},
\end{equation}
where $a$ is the amplitude for the conservative repulsive force,
$x_{ij}=|\vvec{x}_{ij}|$, $\vvec{v_{ij}}=\vvec{v}_{i}-\vvec{v}_{j}$,
$\vvec{v}_{i}$ is the velocity of the $i$th bead. The dissipative
force has an amplitude $\gamma$ and decays with distance according
to the weight function $w^{\mathrm{D}}(x_{ij})$. The amplitude for
the random force is $\sigma$ and the respective weight function is
$w^{\mathrm{R}}(x_{ij})$. $\theta_{ij}$ is the Gaussian random
variable. As was shown by Espa\~{n}ol and Warren \cite{Espanol1995},
to satisfy the detailed balance requirement, the amplitudes and
weight functions for the dissipative and random forces should be
interrelated: $\sigma^2=2\gamma$ (we choose $\gamma=6.75$, $\sigma=\sqrt{2\gamma}=3.67$ here)
and $w^{\mathrm{D}}(x_{ij})=\left[w^{\mathrm{R}}(x_{ij})\right]^2$. Here
we use the weight functions quadratically decaying with a distance:
\begin{equation}
w^{\mathrm{D}}(x_{ij})=\left[w^{\mathrm{R}}(x_{ij})\right]^2
=\left\{
\begin{array}{ll}
(1-x_{ij})^2, & x_{ij} < 1,\\
0, & x_{ij} \geq 1.
\end{array}
\right.
\end{equation}
Because each bead represents an atomic conglomerate rather than a single atom, the interaction potentials between beads
are effective, i.e. depend on the density and the temperature. Groot and Warren \cite{Groot1997}, back in 1997, suggested parametrization approach based on matching the compressibility of the simulated model system to that of water at normal conditions. This yields the amplitude for the conservative repulsive force $a=25$ at reduced number density $\rho^{*} = N/V=3$ of beads ($N$ is the total number of beads in the simulation box of volume $V$). These particular parameters have been used in a numerous studies by many authors since then, including some of ours \cite{Ilnytskyi2007,Kalyuzhnyi2016,Kalyuzhnyi2019} and prove their reliability. Therefore, here we use the same set of parameters.
We consider the case of infinitely diluted pom-pom polymer. To avoid possible artifacts of  the periodic boundary conditions the dimensions of a cubic simulation box was chosen to be at least $1.75(3N_f-1)^0.59$, where $N_f$ is number of beads in a single arm of pom-pom polymer.

The set of  universal size ratios of a pom-pom polymer, as introduced in Sec.2, is obtained from the simulations in the $d=3$ space.
As already pointed out in Sec.2, the pom-pom architecture can be split for the analysis into three constituents: the left- and the right-hand
poms and the backbone. In this study the case of a symmetric pom-pom is considered only, $f_1=f_2=f$ (the total number of branches is, therefore, $F=2f+1$). The non-symmetric case will be the subject of a separate study. The length of each branch, $N_f$, is the same for all branches and we consider two cases: $N_f = 8$ and $N_f=16$. Results for both cases are found to agree well within respective numerical errors, following the previous findings that the coarse-grained chain in DPD approach reaches scaling regime at the length of about $10$ monomers \cite{Ilnytskyi2007, Kalyuzhnyi2016}. In this case the universal shape characteristics became independent on the chain length \cite{Kalyuzhnyi2016}. The case of athermal solvent is assumed, where the amplitude $a$ of a repulsive force in Eq.~(\ref{FC})
is set equal to $25$ for all combinations of the interacting beads: polymer-polymer, solvent-solvent and polymer-solvent.

The time-step is $0.04$ in reduced units. In most cases about $4\cdot10^6$ DPD steps are performed at each number of branches $F$. The first $4\cdot10^5$ steps are allowed for the system equilibration and are skipped from the analysis. The error estimates are made by splitting the duration of a productive run into four equal intervals and evaluation the partial averages $A_i$, $i=1-4$ for a given property $A$ in each of them. The final average is $\langle A \rangle=\frac{1}{4}\displaystyle{\sum_{i=1}^{4}A_i}$, whereas the conservative estimate for the standard error is assumed as $e(A)=\left(\frac{1}{4}\displaystyle{\sum_{i=1}^{4}(A_i-\langle A \rangle)^2}\right)^{\frac{1}{2}}$.

\subsection{Simulation results}

The cases considered in the simulation part of this study include the pom-pom polymer with the
total number of branches ranging from $F=1$ to $17$. The case $F=1$ represents rather formal limit
with no poms and only a backbone of $N_f$ beads being present, it turned to be useful for extrapolation of some results.
The other limit case, $F=3$, reduces pom-pom polymer to a linear chain of a total length of $3N_f$ beads, as far as
both poms and a backbone are linear chains of $N_f$ beads. The molecular topology of the $F=5$ pom-pom is often termed
as a H-polymer \cite{Bishop93a,Radke96,Zweier09}. We should remark that the task to match the simulation results, obtained on a model system of a moderate size and with coarse-grained potentials, to the experimental findings obtained on real systems is nontrivial. Given the range of existing experimental, theoretical and simulation approaches that use different length scales and interaction potentials, we see the reason of comparing universal dimensionless size ratios instead.

\begin{figure}[h!]
\begin{center}
\includegraphics[width=5cm,angle=270]{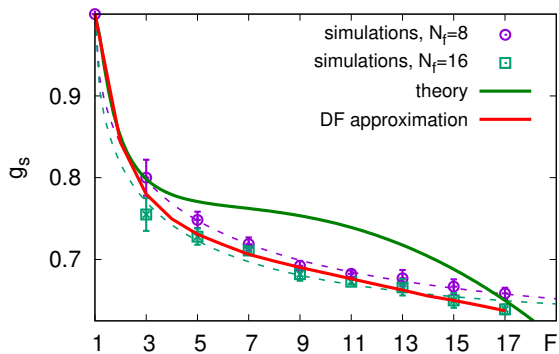}
\caption {\label{fig_gs} Size ratio $g_{s}$ as a function of the number of branches $F$. Simulation data of the pom-pom structure with both $N_f=8$ and $N_f=16$ are shown and indicated in the figure. Dashed lines show their numeric fits: $g_{s}\approx 1 - 0.39(1-e^{-0.52\sqrt{F-1}})$and  $g_{s}\approx 1 - 0.38(1-e^{-0.66\sqrt{F-1}})$, respectively. Solid green curve shows theoretical results in the first order of coupling constant given by Eq.~(\ref{gs}), red solid line -- the same as latter, but with the use of the Douglas-Freed approximation\cite{Douglas84}.}
\end{center}
\end{figure}
Let us start from the results obtained for the $g_s$ universal size ratio as defined in Eq.~(\ref{gs}). It measures
the extent to which the star polymer with $F$ branches is more compact in its shape than the symmetric pom-pom polymer
with the same total number of branches $F$. The results are visualized in Fig.~\ref{fig_gs}. First, note that the simulation data
obtained for $N_f=8$ and $N_f=16$ coincide within respective numeric errors and their numeric fits, shown in the caption of this figure, contain very close coefficient. Another note is the presence of a formal limit $g_s=1$ at $F=1$, and at $F=3$ the meaning of $g_s(3)$ coincides with that for the size ratio $g(3)$ between the star and a linear chain, as defined in Ref.~\citenum{Kalyuzhnyi2019}. Both values are found to be the same, $g_s(3)\approx g(3)\approx 0.8$, see Fig.~\ref{fig_gs} of this study and Fig.~2 in Ref.~\citenum{Kalyuzhnyi2019}. With the increase of $F$, both theoretical result
and the simulation data show the decrease of $g_s$ in Fig.~\ref{fig_gs}, indicating a more compact size of a star
comparing to an equivalent pom-pom, which is expected due to their respective molecular architectures. Both methods give quite close results at $F=1-5$ and $F=15-17$, but at other values of $F$ the discrepancy between the results still does not exceed $7\%$. Let us note that the Douglas-Freed approximation for the theoretical results agree extremely well with the simulation data in the whole interval of values $1<F<17$ being considered. This approximation maps to some extent
higher order terms of perturbation theory indicating that these are needed indeed to achieve better match between the theory and the simulation data.

 \begin{figure}[h!]
\begin{center}
\includegraphics[width=5cm,angle=270]{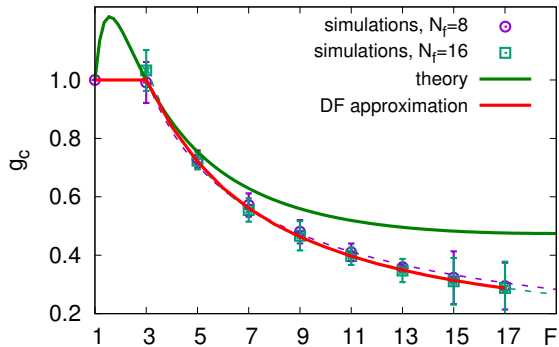}
\caption {\label{fig_gc} The same as in Fig.~\ref{fig_gs} but for the size ratio $g_{c}$ given by Eq.~(\ref{gc}). Numeric fits are: $g_{c}\approx 2.14 F^{-0.69}$ and $g_{c}\approx 2.37 F^{-0.74}$ for the $N_f=8$ and $N=16$ cases, respectively.}
\end{center}
\end{figure}
The $g_c$ universal size ratio is defined in Eq.~(\ref{gc}). It characterizes compactness of the pom-pom polymer with $F$ branches
comparing to the linear chain of the length $N_f\,F$ beads with the increase of $F$. The results are shown in Fig.~\ref{fig_gc} and, similar to the $g_s$ ratio, the simulation data for the $N_f=8$ and $16$ cases are extremely close. One should note again
that at $F=1$ and $3$ the ratio $g_c=1$ by its definition. At $F>3$ the dependence can be fitted well by a
power law indicated in the figure caption. We observe much steeper decay here than for the $g_s$. It is to be expected as both a star
and a pom-pom are much more compact compared to the equivalent molecular mass linear chain, see Fig.~2 in Ref.~\citenum{Kalyuzhnyi2019}
and Fig.~\ref{fig_gc} in a current study. Hence, $g_s$ provides the ratio between the squared gyration radii of two relatively
compact objects and, hence at $F>3$ it is found in a rather narrow interval of $0.65<g_s<0.8$. Let us also note, that the results for $g_c$ obtained
theoretically and via computer simulations match each other very well for $F\leq 7$. It is also worth mentioning that the simulation give stronger
effect of pom-pom ``compactization'' with respect to the linear chain than predicted by a theory, in contrary to all other ratios considered below. Again,
the Douglas-Freed approximation agree very well with the simulation data at $1<F<17$.

\begin{figure}[h!]
\begin{center}
\includegraphics[width=5cm,,angle=270]{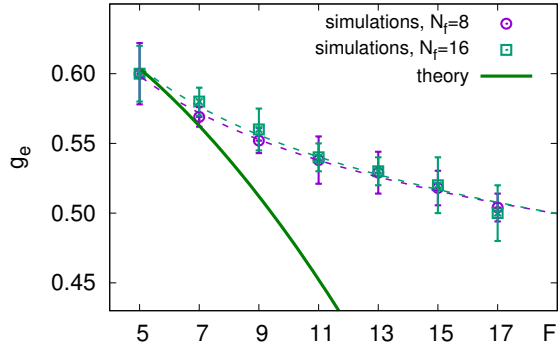}
\caption {\label{fig_ge} The same as in Fig.~\ref{fig_gs} but for the ratio asymmetry ratio $g_{e}$ given by Eq.~(\ref{ge}). Numeric fits are: $g_{e}\approx 0.74 F^{-0.14}$ and $g_{e}\approx 0.76 F^{-0.14}$ for the $N_f=8$ and $N=16$ cases, respectively.}
\end{center}
\end{figure}
Analysis of the shape characteristics of a pom-pom molecule as a single entity is completed by considering the asymmetry ratio $g_e$ defined in Eq.~(\ref{ge}).
It is based on evaluation of the average distances between the ends of the branches. More precisely, it is the ratio between the average intra-pom
end-to-end distance and its inter-pom counterpart. This ratio provides an average pom-pom anisotropy measured perpendicularly and parallel
to the line connecting two of its branching points. By definition, it is defined at $F\geq 5$ only. The plot for $g_e$, shown in Fig.~\ref{fig_ge}, shows an excellent agreement between the theory
and computer simulations at $F\leq 7$. With the increase of $F$, the theory predicts more rapid decrease of $g_e$, in contrary to the case of $g_c$.
The simulation data for $N_f=8$ and $16$ coincide again and indicate rather moderate rate for the $g_e$ decrease with $F$, fitted well by the power law expressions provided in the figure caption.

\begin{figure}[h!]
\begin{center}
\includegraphics[width=5cm,angle=270]{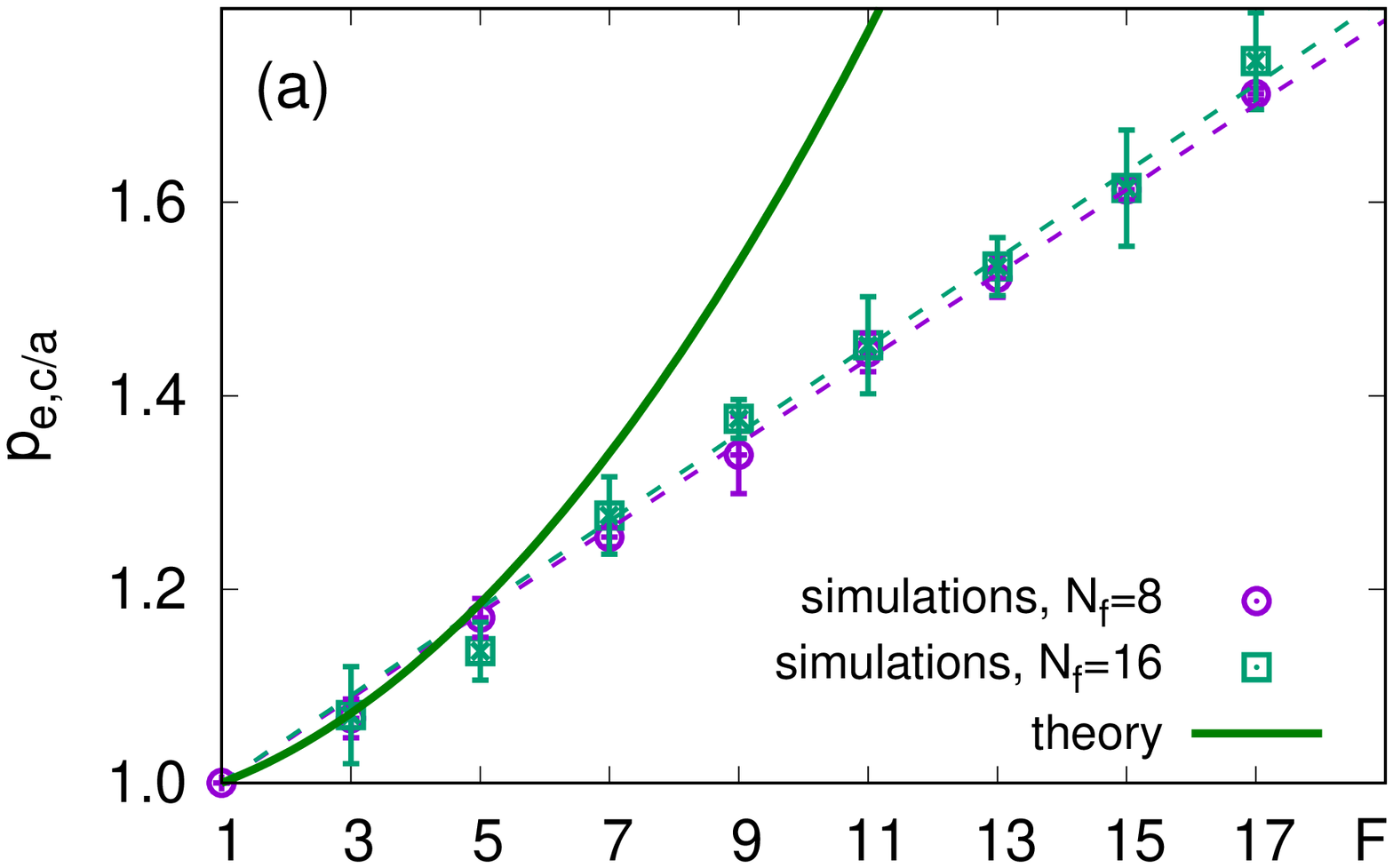}
\includegraphics[width=5cm,angle=270]{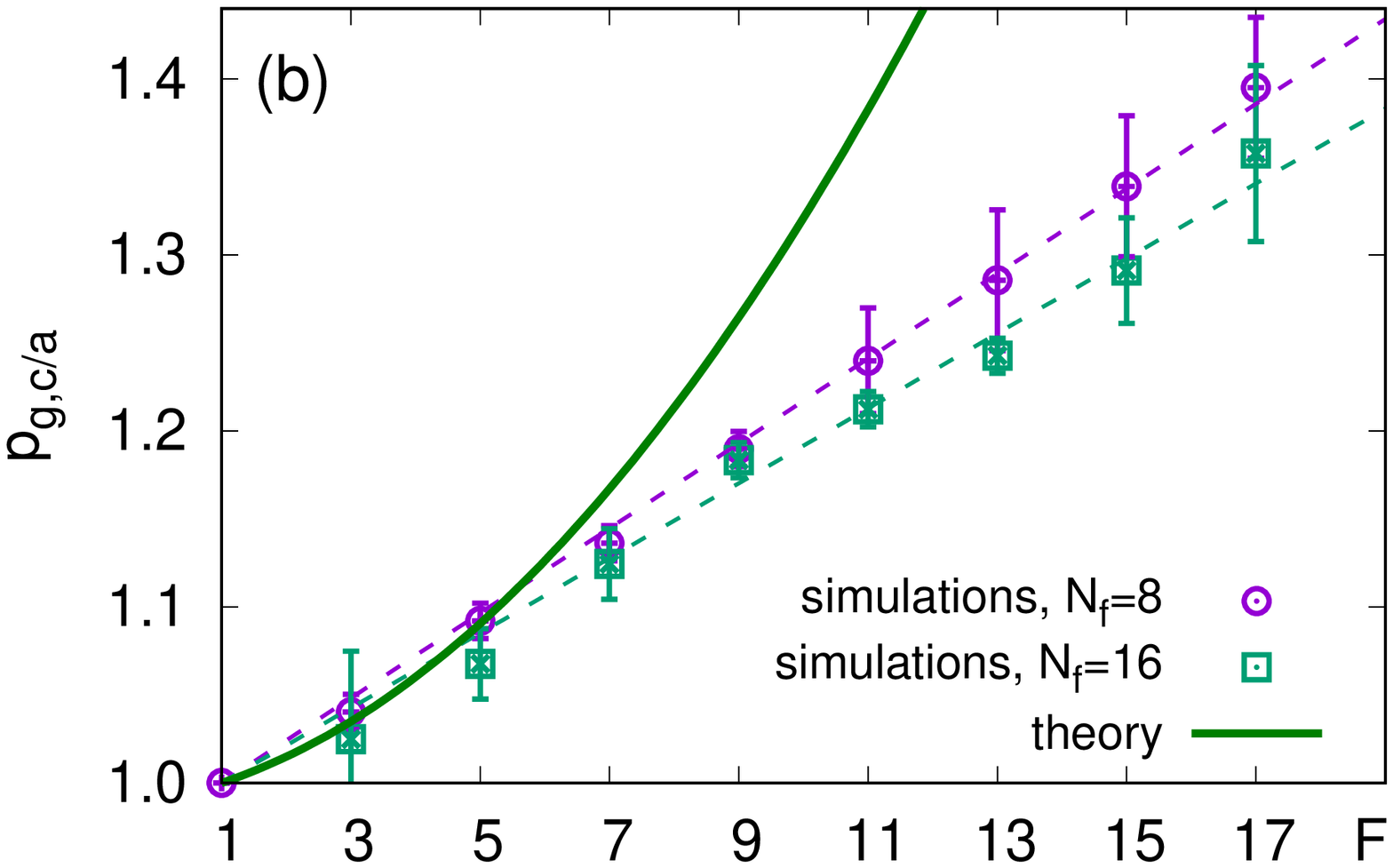}
\caption {\label{fig_peca_pgca} (a) The stretch ratio $p_{e,c/a}$ as a function of number of branches $F$, numeric fits are: $p_{e,c/a}\approx 1 + 0.044(F-1)$ and $p_{e,c/a}\approx 1 + 0.045(F-1)$ for the $N_f=8$ and $N_f=16$ cases, respectively. (b) The same for the swelling ratio $p_{g,c/a}$, numeric fits are: $p_{g,c/a}\approx 1 + 0.024(F-1)$ and $p_{g,c/a}\approx 1 + 0.021(F-1)$ for the $N_f=8$ and $N_f=16$ cases, respectively. Theoretical results are given by Eqs.~(\ref{peca_res}) and (\ref{pgca_res}), respectively.}
\end{center}
\end{figure}
Let us consider the stretch and swelling ratios $p_{e,c/a}$ and $p_{g,c/a}$, defined in Eq. (~\ref{pca}), now. The former quantifies the squared end-to-end distance of a backbone compared to that for pom branches, the latter -- the same for the squared gyration radius. Both are plotted in Fig.~\ref{fig_peca_pgca}. We will remark these ratios are both equal to $1$ only at the formal limit $F=1$. At $F=3$, when the pom-pom reduces itself to a three-part linear chain, the magnitudes of both ratios are higher than one, i.e. a backbone is on average more stretched than pom branches. This is explained by clamping of the backbone ends to the ends of the pom branches and, therefore, restricting the ability of a backbone to adopt a compact coil form. Both $p_{e,c/a}$ and $p_{g,c/a}$ ratios increase with the growth of $F$. This is attributed to the fact that both poms are getting gradually denser (see the effect for a single star polymer in Ref.~\citenum{Kalyuzhnyi2019}) and, as a consequence, stretch a backbone that links them. In this case, lowering of the internal energy contribution to the free energy compensate the decrease of a conformation entropy of the stretched backbone. Simulation data for $N_f=8$ and $16$ agree well within the numerical accuracy of simulations given by the error bars. Let us also note that the theoretical results given by Eqs.~ (\ref{peca_res}) and (\ref{pgca_res}) predict stronger and a non-linear growth of both ratios with $F$. The simulation data show the dependence close to linear, numeric fits are given in the caption of Fig.~\ref{fig_peca_pgca}. We see that the rate of the $p_{e,c/a}$ growth with $F$ is almost twice higher than that for $p_{g,c/a}$.

\begin{figure}[h!]
\begin{center}
\includegraphics[width=5cm,angle=270]{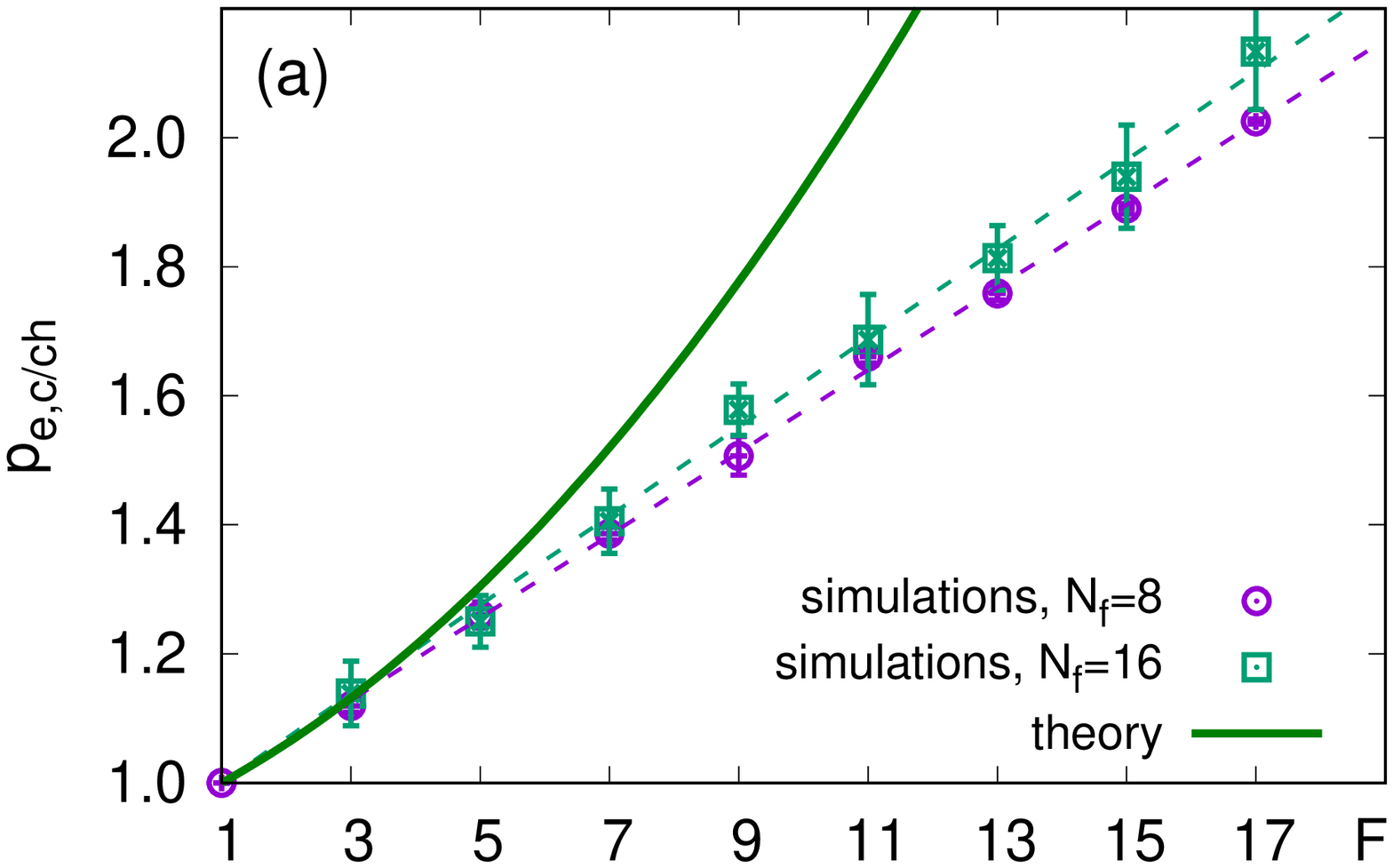}
\includegraphics[width=5cm,angle=270]{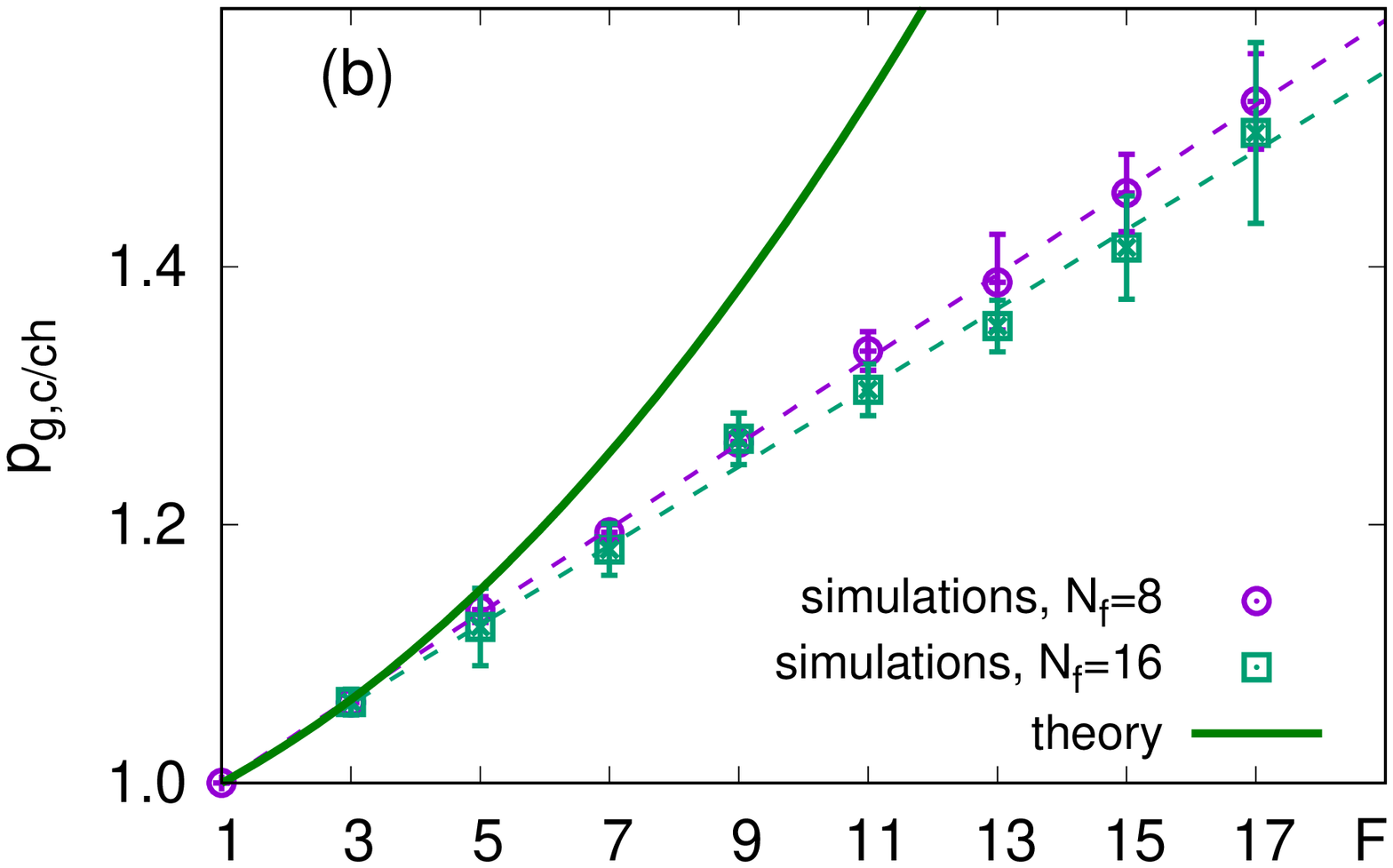}
\caption {\label{fig_pecch_pgcch} (a) The stretch ratio $p_{e,c/ch}$ as a function of number of branches $F$, numeric fits are: $p_{e,c/ch}\approx 1 + 0.064(F-1)$ and $p_{e,c/ch}\approx 1 + 0.069(F-1)$ for the $N_f=8$ and $N_f=16$ cases, respectively. (b) The same for the swelling ratio $p_{g,c/ch}$, numeric fits are: $p_{g,c/ch}\approx 1 + 0.033(F-1)$ and $p_{g,c/ch}\approx 1 + 0.031(F-1)$ for the $N_f=8$ and $N_f=16$ cases, respectively. Theoretical results are given by Eqs.~(\ref{pecch_res}) and (\ref{pgcch_res}), respectively.}
\end{center}
\end{figure}
The second pair of  stretch and swelling ratios $p_{e,c/ch}$ and $p_{g,c/ch}$ characterize the backbone end-to-end distance and the squared gyration radius compared to their counterparts of a single free chain with the same number of monomers $N_f$, see definitions (\ref{pccch}). Their dependencies on $F$ are shown in Fig.~\ref{fig_peca_pgca} and again the simulation data for $N_f=8$ and $16$ agree well within the estimated numerical accuracy. The theoretical results given by Eqs.~(\ref{pecch_res}) and (\ref{pgcch_res}) agree very well with the simulation data at $F\leq 7$ but predict stronger non-linear growth with $F$ comparing to the simulation data at higher $F$. Simulation data again is fitted well by linear forms provided in the caption of Fig.~\ref{fig_peca_pgca}. By comparing respective fitting expressions, one may conclude that the effect of the backbone stretch with the increase of $F$ is much stronger than the stretch of the pom branches, when compared to the free linear chain.

\begin{figure}[h!]
\begin{center}
\includegraphics[width=50mm,,angle=270]{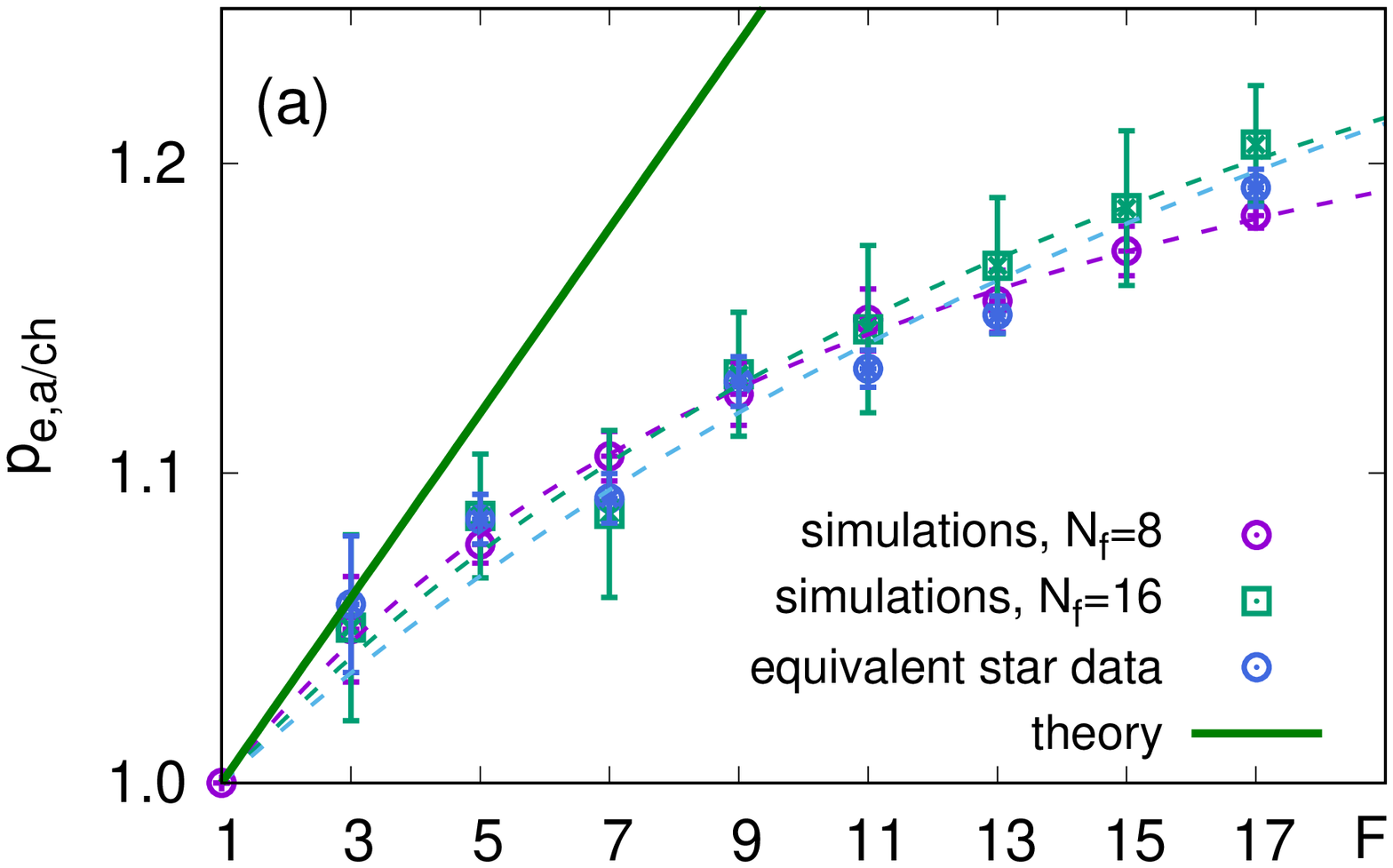}
\includegraphics[width=50mm,,angle=270]{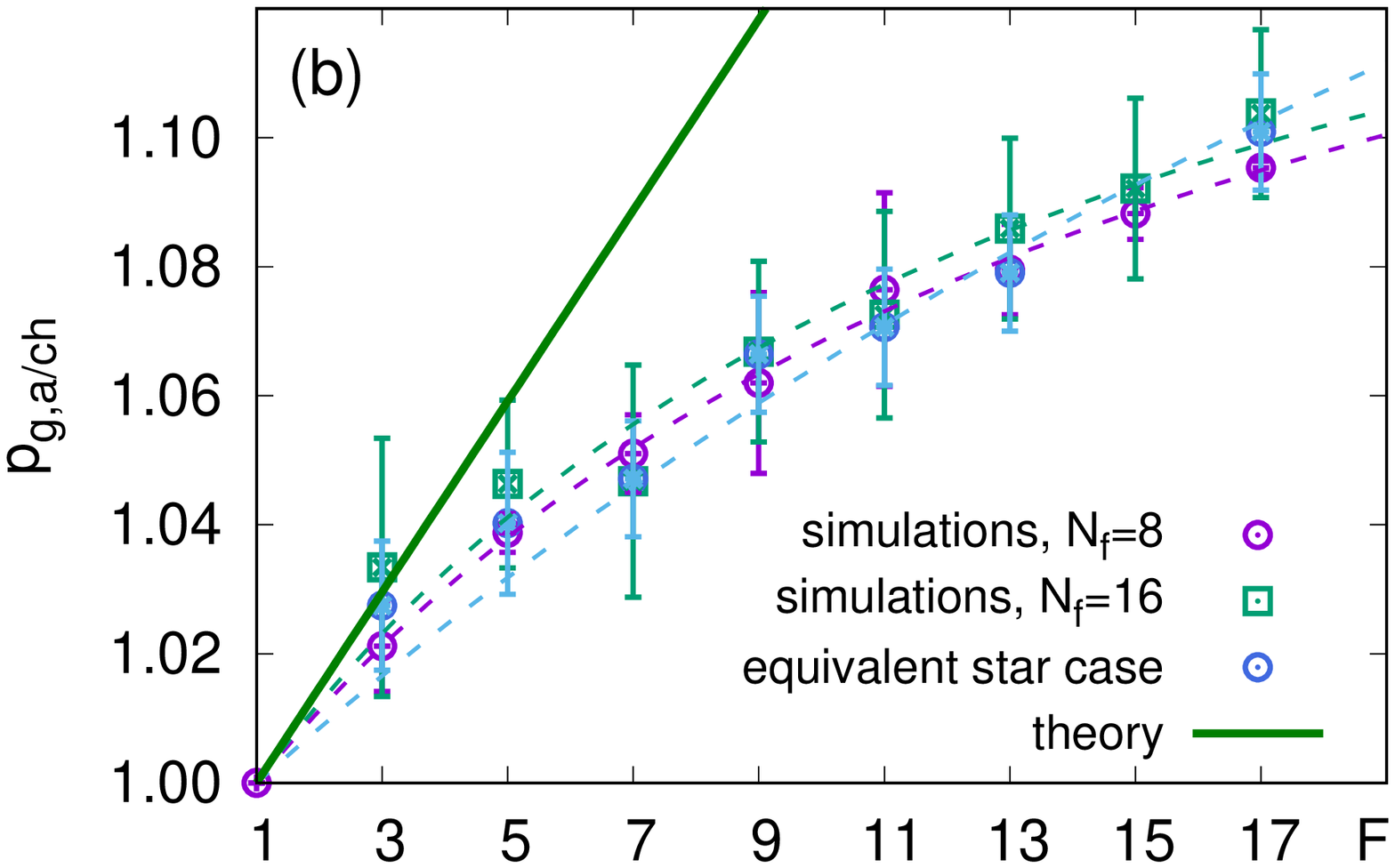}
\caption {\label{fig_peach_pgach} (a) The stretch ratio $p_{e,a/ch}$ as a function of number of branches $F$, simulation data for the equivalent star case (see explanation in the text) is also shown and indicated. Numeric fits are: $p_{e,a/ch}\approx \frac{1+0.109(F-1)}{1+0.083(F-1)}$, $p_{e,a/ch}\approx \frac{1+0.070(F-1)}{1+0.047(F-1)}$ and $p_{e,a/ch}\approx \frac{1+0.052(F-1)}{1+0.033(F-1)}$ for the $N_f=8$, $N_f=16$ and equivalent star cases, respectively. (b) The same for the swelling ratio $p_{g,a/ch}$, numeric fits are: $p_{g,a/ch}\approx \frac{1+0.075(F-1)}{1+0.063(F-1)}$, $p_{g,a/ch}\approx \frac{1+0.084(F-1)}{1+0.071(F-1)}$ and $p_{g,a/ch}\approx \frac{1+0.031(F-1)}{1+0.022(F-1)}$ for the $N_f=8$, $N_f=16$ and equivalent star cases, respectively. Theoretical results are given by Eqs.~(\ref{peach_res}) and (\ref{pgach_res}), respectively.}
\end{center}
\end{figure}
This, indeed is found in the direct evaluation of the related ratios, $p_{e,a/ch}$ and $p_{g,a/ch}$ defined in Eq.~(\ref{pach}). Both are shown in Fig.~\ref{fig_peach_pgach} and one can see that their magnitudes does not exceed $1.2$ and $1.1$, respectively at the maximum number of branches $F=17$ being considered. These are $1.5-2$ times lower than the maximum values for $p_{e,c/ch}$ and $p_{g,c/ch}$ in Fig.~\ref{fig_pecch_pgcch}. Here one can see that the theoretical result given by Eqs.~(\ref{peach_res}) and (\ref{pgach_res}) increases linearly with $F$, but the overal shape of the simulation data dependencies display bending at higher $F$. We tried several fitting expressions including power-law and logarithmic functions, but to less success than the rational fractions as indicated in the caption of Fig.~\ref{fig_peach_pgach}. We also found good agreement between the $N_f=8$ and $N_f=16$ cases. One should also note that $p_{e,a/ch}$ and $p_{g,a/ch}$ ratios for the poms branches quantify the same effect of their stretch with the increase of $F$,  as the $p_e(f)$ and $p_g(f)$ defined in Ref.~\citenum{Kalyuzhnyi2019} for a single star with $f$ branches. We compared effects in both cases taking into account that the single star-like polymer representing each pom in a pom-pom architecture has $f'=(F-1)/2+1$ branches, assuming that the backbone is shared between the two pom as an additional branch for each pom. The simulation data for a single star with $f'$ branches should be matched then with the data obtained for the pom-pom polymer with $F=2f'-1$ total branches. This is confirmed in Fig.~\ref{fig_peach_pgach}, where the pom-pom and the equivalent single star polymer with $f$ branches of $N_f=8$ monomers each are compared and found to match really well within the accuracy of simulations. Hence, one may conclude that both poms, in terms of the stretch of their branches, behave almost exactly the same as free stars.

Based on the error estimates indicated in each of Figs.~\ref{fig_gs}-\ref{fig_peach_pgach}, one can note that the dissipative particle dynamics simulations provide variable but reasonable accuracy for all shape characteristics being introduced and evaluated in this study.

It is important to note at this point that both methods have their strong sides as well as limitations. On one hand, the definition of the polymer topology in the continuous chain model through the partition function allows to describe relatively easy a wide range of topologies in one loop approximation, however the calculations of the higher orders of perturbation theory, that are necessary for a better agreement, become extremely time consuming as well as requiring a much more complicated mathematics. Also the use of perturbation theory limits the model to the cases with low values of branching parameter $F$, thus leading to some discrepancies as compared with the simulations.

On the other hand, a DPD simulations where conducted with a soft potential, in contrast to a hard delta-potential in the theory. It is known that the use of such a potential provides correct scaling behaviour for the linear chain {see ref.~\cite{Ilnytskyi2007} and references therein}, but one is limited in terms of a system size and simulation times to make the expenditure of computer time feasible.

Although, both theory and simulations are contributing to the discrepancies between their respective findings, the agreement found between simulations and the theory at moderate $F$ is very encouraging. On a top of that, an excellent agreement between the theoretical results using the Douglas-Freed approximation and the simulations in a broad interval of $F$ was a complete surprise which calls for the further investigation in this direction.

\section{Conclusion}

In this paper we combine theoretical studies and computer simulations to evaluate the set of universal size ratios of the symmetric pom-pom polymers. The study is motivated by the role of the macromolecular shape in a range of applications. Theoretical approach is centered around direct renormalization of the Edwards continuous chain model and involves evaluation of the pom-pom size ratios characterizing both the whole molecule and its individual branches in the first order in $\epsilon=4-d$ expansion. Computer simulations employ the dissipative particle dynamics mesoscopic simulation technique for the pom-pom polymer in a good solvent with two cases of the branch length of $8$ and $16$ beads being considered. We found that the universal ratios for the shape characteristics obtained at these two branch lengths do agree well within their respective numerical errors.

The stretch and ``compactization'' of the pom-pom polymer as a whole and of its individual branches that belong to poms and to its backbone, with the increase of the total number of branches $F$, are characterised by $9$ universal ratios. Some of them are introduced in this study for the first time. We should note that the theoretical and the computation models do differ essentially in their respective description of the polymer structure on both energy and length scales. Despite that, we found an excellent agreement between the theoretical and simulation results for almost all universal ratio of size characteristics at moderate values of $F\leq 7$. On a top of that, truly remarkable agreement between the simulations and the theoretical results for the two ratios, for which one can use of the Douglas-Freed approximation, in the whole interval of $F$ being considered, was a complete surprise. We see in this the true manifesto to the universality concept in relation to the polymer properties. Still, the discrepancies between the at higher $F$ may be attributed to the need to evaluate higher order approximations in the theory, but may also be the consequence of the soft nature of the interaction potentials used in the dissipative particles dynamics.
}

The work may be extended in several directions, e.g. by: performing higher order calculations in the theory; considering the shape of the asymmetric pom-pom polymers; studying aggregation of amphiphilic pom-pom polymers, considering their shape in a Poiseuille flow, etc.

\section*{Appendix}
Here, we provide a list of the expressions for the Gyration radii in general space dimension.

Gyration radius of a linear chain:
  \begin{eqnarray}
  &&\langle R_{g,chain}^2\rangle=\frac{dL}{6}\left(1+u_0\frac{2d^4-56d^3+376d^2-544d}{d(d-6)(d-8)(d-10)(d-2)(d-4)}\right).
  \end{eqnarray}

Gyration radius of a star polymer with $F$ branches:
  \begin{eqnarray}
  &&\langle R_{g,star}^2\rangle=\frac{dL(3F-2)}{6F^2}\left(1+u_0\times\nonumber\right.\\
  &&\left.\left(\frac{2^{4-d/2}((F-1)dF(d-2)(d^2-26d+136)+1920(F-2))}{d(d-6)(d-8)(d-10)(d-2)(d-4)(3F-2)}\right.\right.\nonumber\\
  &&-\frac{2(F-2)(4F(d-10)(d-12)(d^2-6d+32)-3d^4)}{d(d-6)(d-8)(d-10)(d-2)(d-4)(3F-2)}\nonumber\\
  &&\left.\left.-\frac{2(F-2)(84d^3-948d^2+5424d-15360)}{d(d-6)(d-8)(d-10)(d-2)(d-4)(3F-2)}\right)\right).
  \end{eqnarray}

Gyration radius of a pom-pom polymer:
  \begin{eqnarray}
  &&\langle R_{g,pom-pom}^2\rangle=\frac{dL{(18f^2+8f+1)}}{6(2f+1)^2}\left(1+u_0\times\nonumber\right.\\
  &&\frac{2(18f^2+8f+1)^{-1}}{d(d-6)(d-8)(d-10)(d-2)(d-4)}\times\nonumber\\
  &&\left(\,3^{2-d/2}f^2(864d(2f+1)(d-10)-(7680(f+5))(f-1))\right.\nonumber\\
  &&-d(2f+1)(2f+7)(d-14)(d^2-14d+88)\nonumber\\
  &&+2^{3-d/2}(6d^{10}f^3-3d^{10}f^2-348d^9f^3-3d^{10}f+174d^9f^2\nonumber\\
  &&+8208d^8f^3+174d^9f-4104d^8f^2-103392d^7f^3-4104d^8f\nonumber\\
  &&+51696d^7f^2+772704d^6f^3+51696d^7f-409392d^6f^2\nonumber\\
  &&-3690432d^5f^3+6d^4f^4-374832d^6f+2536416d^5f^2\nonumber\\
  &&+12444676d^4f^3-168d^3f^4+11520d^6+1499616d^5f\nonumber\\
  &&-14055948d^4f^2-32222320d^3f^3+1704d^2f^4-345600d^5\nonumber\\
  &&-2305542d^4f+57583408d^3f^2+56771504d^2f^3-7392df^4\nonumber\\
  &&+3916800d^4-4624760d^3f-129395664d^2f^2-44241344df^3\nonumber\\
  &&+11520f^4-20736000d^3+22117272d^2f+110606144df^2+23040f^3\nonumber\\
  &&+50503680d^2-22115104df-38400f^2-44236800d+3840f)\nonumber\\
  &&-12d^{10}f^4+21d^{10}f^3+696d^9f^4+15d^{10}f^2-1218d^9f^3\nonumber\\
  &&-17568d^8f^4-18d^{10}f-870d^9f^2+31032d^8f^3\nonumber\\
  &&+255168d^7f^4+1044d^9f+21672d^8f^2-458640d^7f^3-2374848d^6f^4\nonumber\\
  &&-26928d^8f-306864d^7f^2+4369104d^6f^3+14845824d^5f^4\nonumber\\
  &&+406944d^7f+2755440d^6f^2-28019232d^5f^3-62656524d^4f^4\nonumber\\
  &&-3988512d^6f-16518240d^5f^2+121049120d^4f^3\nonumber\\
  &&+170944848d^3f^4+26346816d^5f+66920439d^4f^2-336146816d^3f^3\nonumber\\
  &&-267636048d^2f^4-116785160d^4f-176687348d^3f^2+532141120d^2f^3\nonumber\\
  &&+176961984df^4-2d^4+330403040d^3f+270763044d^2f^2-353938816df^3\nonumber\\
  &&-23040f^4+48d^3-529000672d^2f-176930608df^2+92160f^3\nonumber\\
  &&\left.\left.-376d^2+353905792df-38400f^2+960d-30720f\right)\right),
  \end{eqnarray}
with $f$ being the number of side branches on one branching center of the pom-pom, making that a total amount of branches $F=2f+1$

In general, these expressions can be presented as $\langle R_{g,x}^2\rangle=\langle R_{g,x}^2\rangle_0 (1+u_0X(F,d))$, and the coefficients of the two parameter model (\ref{df}) can be determined from the expression:
\begin{equation}
C(F)=F^{-d/2}X(F,d=3).
\end{equation}

\bibliographystyle{unsrt}
\bibliography{Pom-Pom}

\newpage

\end{document}